\documentclass{PoS}

\title{In-medium Properties of Hadrons}

\ShortTitle{In-medium Properties of Hadrons}

\author{\speaker{Mariana Nanova}%
        \thanks{on behalf of the CBELSA/TAPS Collaboration}\\
        II. Phys. Institut, JLU - Giessen\\
        E-mail: \email{Mariana.Nanova@exp2.physik.uni-giessen.de}}

\abstract{The changes of hadron properties within strongly interacting matter provide a link between experimental observables and Quantum Chromodynamics (QCD) in the non-pertubative sector. The sensitivity of various observables to in-medium modifications of hadrons is discussed. The experimental approaches to study in-medium properties, namely  transparency ratio measurement, lineshape analysis, measurement of momentum distributions or excitation functions, are presented based on discussing recent results from the CBELSA/TAPS and CB/TAPS@MAMI experiments.}

\FullConference{50th International Winter Meeting on Nuclear Physics - Bormio2012,\\
		23-27 January 2012\\
		Bormio, Italy}

\begin{document}

\section{Introduction}
\begin{figure*} 
 \resizebox{0.8\textwidth}{!}{
     \includegraphics[height=.1\textheight]{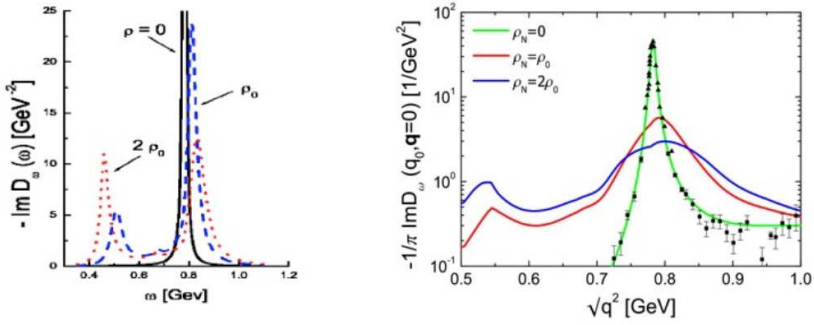}}\\
 \caption{Left: The $\omega$ spectral functions of the free $\omega$ meson (black curve), for normal density $\rho_{0} = 0.16 \ fm^{-3}$ (blue curve) and for $\rho = 2 \cdot \rho_{0}$ \cite{lutz}. Right: The $\omega$ spectral function for an $\omega$ at rest in the nuclear medium. The data and the solid curve decsribe the spectral function of the free $\omega$ meson. The other curves show the spectral function for normal density $\rho_{0} = 0.16 \ fm^{-3}$  and $\rho = 2 \cdot \rho_{0}$, respectively \cite {muhlich}.} \label{fig:theo} 
\end{figure*}

The modification of hadron properties in strongly interacting environment is one of the important topics in hadron and heavy-ion physics. The great interest in this field is motivated by the expectation to find evidence for possible restoration of chiral symmetry in a nuclear medium. Several theoretical papers discuss the idea that the chiral symmetry may at least partially restored in a nuclear medium \cite{meissner,brown,hatsuda}. Hadronic models, based on meson-baryon interactions, calculate the in-medium selfenergies of hadrons and their spectral functions. Corresponding calculations have been performed by many theoretical groups. As an example results for the $\omega$-meson are shown  in Fig.~\ref{fig:theo}. It is seen that spectral functions can have complicated structure with an addidional peak at lower masses (around 500 MeV), due to coupling to  different resonances \cite {muhlich,lutz}. The broadening of the $\omega$-meson in the medium has also been pointed out in different models \cite{muhlich}. Fig.~\ref{fig:theo} illustrates that experiments searching for medium modifications will have to be sensitive to broadening, structures or mass shift. The experiments have focused on the light vector mesons $\rho$, $\omega$ and $\Phi$ with decay lengths comparable to nuclear dimensions. 

The properties of $\eta^\prime $ are largely governed by the dynamics of the QCD $U_{A}(1)$ axial vector anomaly ~\cite{weinberg}.
The light pseudoscalar mesons ($\pi$, $K$, $\eta$) are the Nambu-Goldstone bosons associated with the spontaneous breaking of the QCD chiral symmetry. Introducing the current quark masses these mesons together with the heavier $\eta^\prime$(958) meson show a mass spectrum which is believed to be explained by the explicit flavor $SU(3)$ breaking and the axial $U_{A}(1)$ anomaly. Recently, there have been several important developments in the study of the spontaneous breaking of chiral symmetry and its partial restoration at finite density~\cite{costa,jido}. Indirect evidence has been claimed for a dropping $\eta^\prime$ mass in the hot and dense matter formed in ultrarelativistic heavy-ion collisions at RHIC energies \cite{Csoergo}.\\
An experimental approach to learn about the $\eta^\prime N$ interaction and the in-medium properties of the $\eta^\prime$ meson is the study of $\eta^\prime $ photoproduction off nuclei. The in-medium width of the $\eta^\prime $-meson can be extracted from the attenuation of the $\eta^\prime$-meson flux deduced from a measurement of the transparency ratio for a number of nuclei. Unless when removed by inelastic channels the $\eta^\prime$-meson will decay outside of the nucleus because of its long lifetime and thus its in-medium mass is not accessible experimentally. The in-medium width provides information on the strength of the $\eta^\prime N$ interaction, as studied in \cite{oset_ramos}, and it will be instructive to compare this result with in-medium widths obtained for other mesons. Furthermore, knowledge of the $\eta^\prime$ in-medium width is important for the feasibility of observing $\eta^\prime$ - nucleus bound systems theoretically predicted in some models \cite{Nagahiro}. Until recently, $\eta^\prime$ photoproduction had not been much explored, but with new generation experiments results on differential and total cross sections of $\eta^\prime$ photoproduction on the proton ~\cite{crede} and the deuteron ~\cite{Igal} have now become available. \\

\section{Experimental approaches to study in-medium properties of hadrons}
Different methods to study the in-medium properties of hadrons, applicable to any meson lifetime, are now of great interest and will be discussed.  Results on in-medium properties of the $\omega$ and $\eta^{'}$-meson deduced from transparency ratio measurement are shown. We present our investigations of in-medium modifications of the $\omega$-meson and discuss the sensitivity of the lineshape analysis. Momentum distribution of the hadrons and their excitation function also provide information on their in-medium properties.
\subsection{\it Transparency Ratio Measurement}
An independent access to the in-medium width of hadrons is provided by attenuation measurements on nuclei with different mass number. In a nuclear medium, mesons can be removed by inelastic reactions with neighbouring hadrons. Thereby the lifetime of these mesons are shortened and the widths are increased. The absorption of mesons can be extracted from the measurement of the transparency ratio ~\cite{kaskulov, muhlich1}:\\

\begin{equation}
T_A=\frac{\sigma_{\gamma A\to \omega(\eta^\prime) X}}{A \cdot
\sigma_{\gamma N \to \omega(\eta^\prime) X} } \ .
\label{eq:trans}
\end{equation} 

\begin{figure*} 
 \resizebox{0.5\textwidth}{!}{
     \includegraphics[height=.1\textheight]{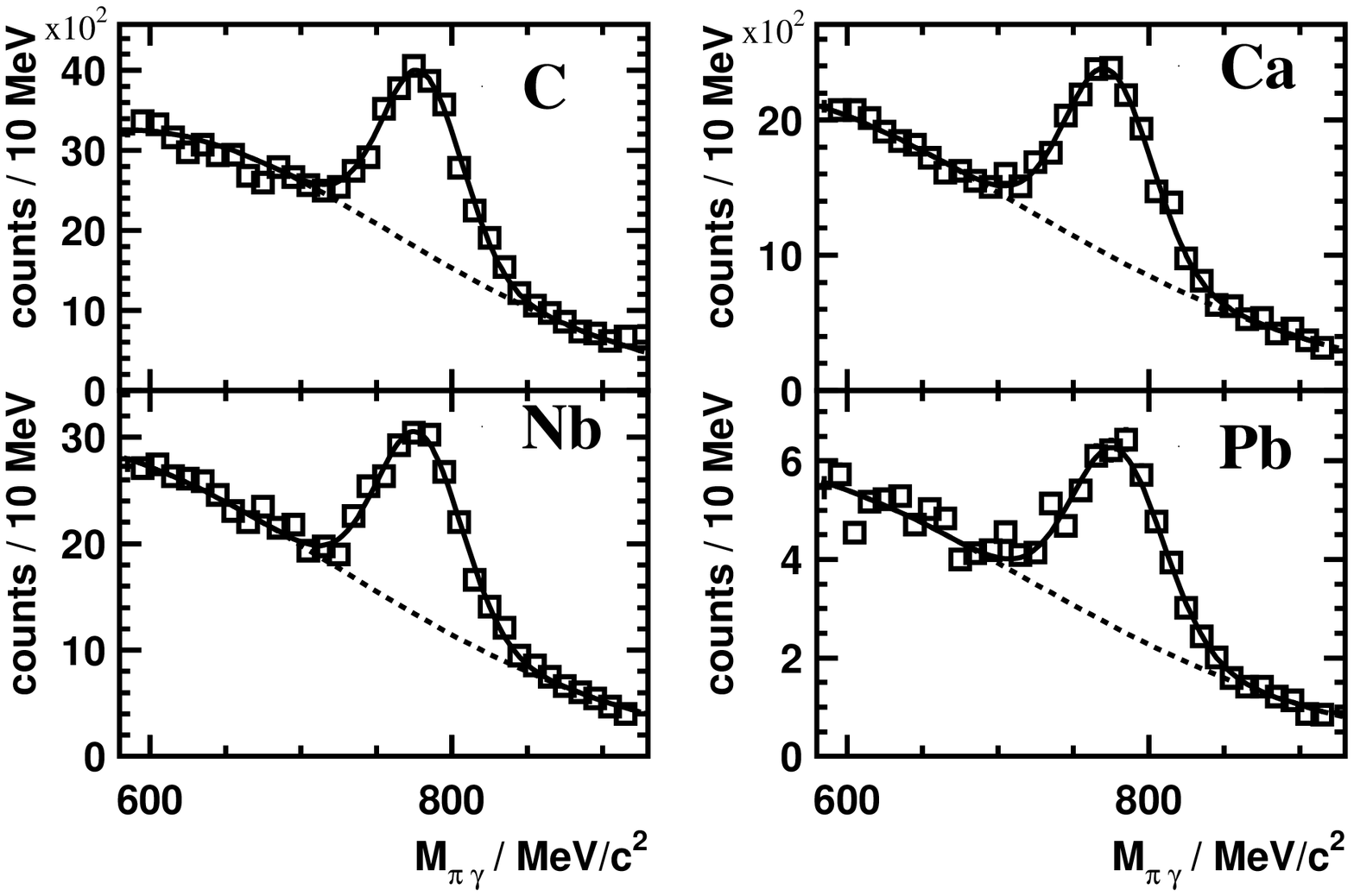}}\\
     
 \resizebox{0.9\textwidth}{!}{    
      \includegraphics[height=.18\textheight]{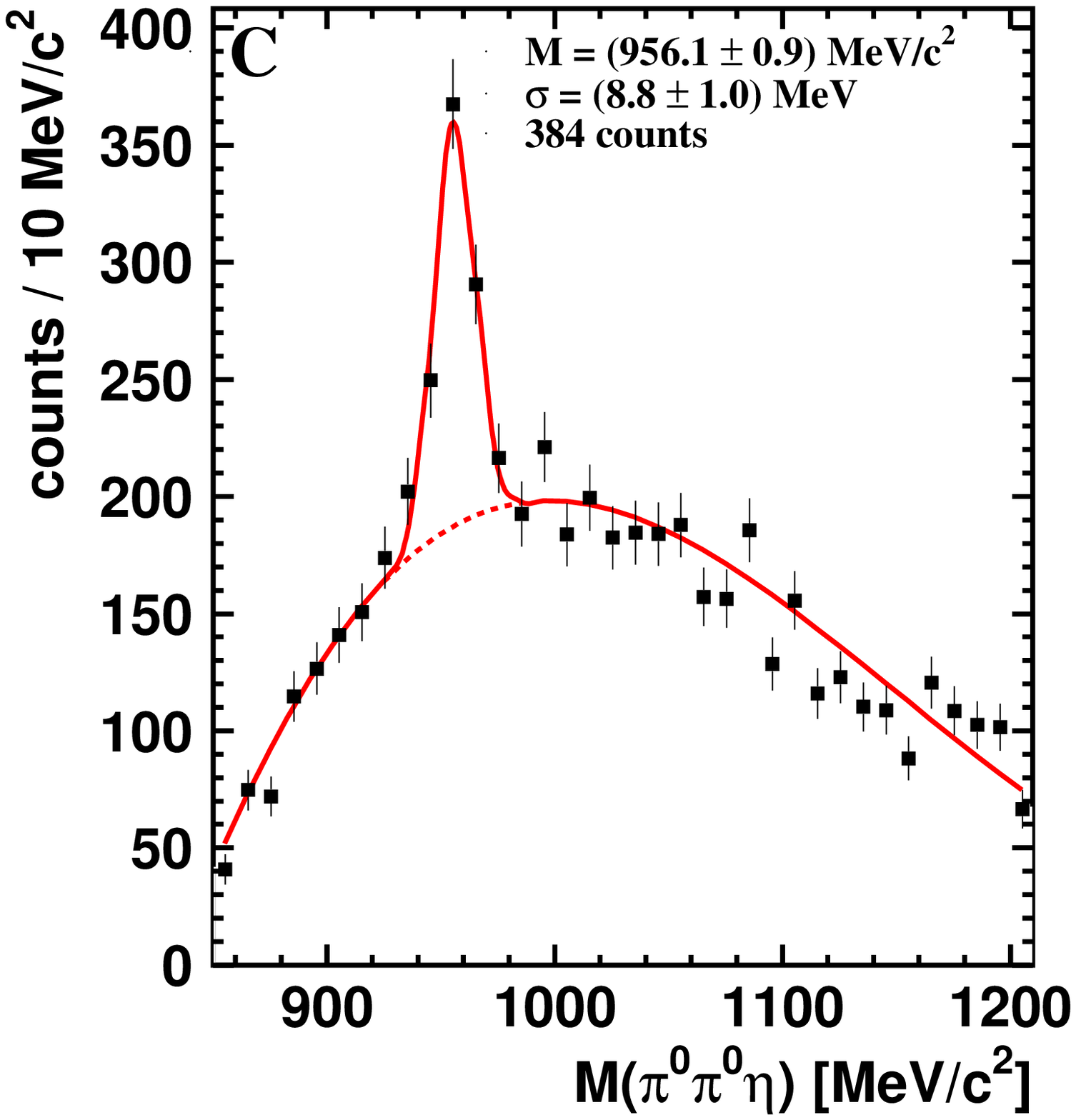} \includegraphics[height=.18\textheight]{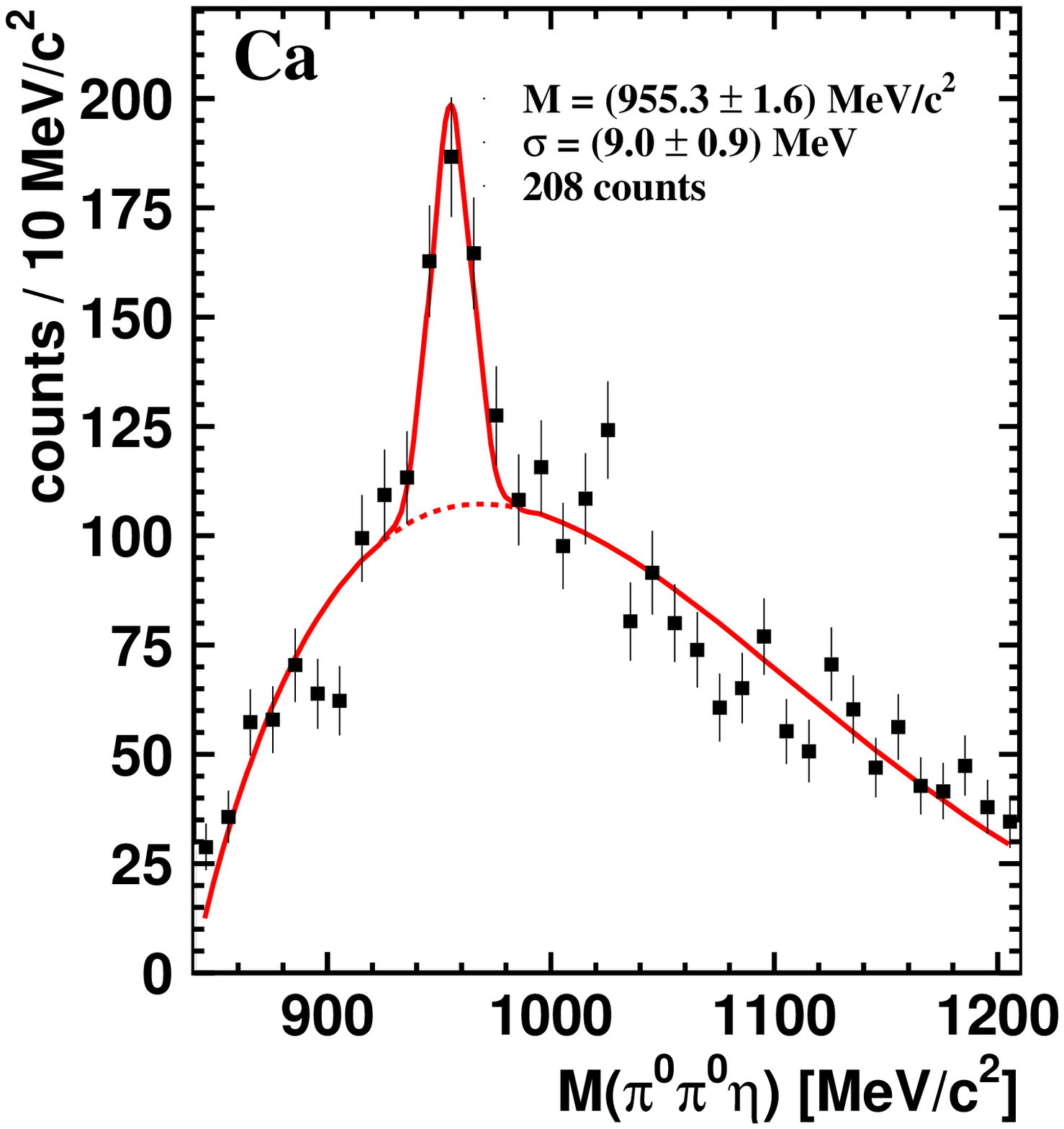} \includegraphics[height=.18\textheight]{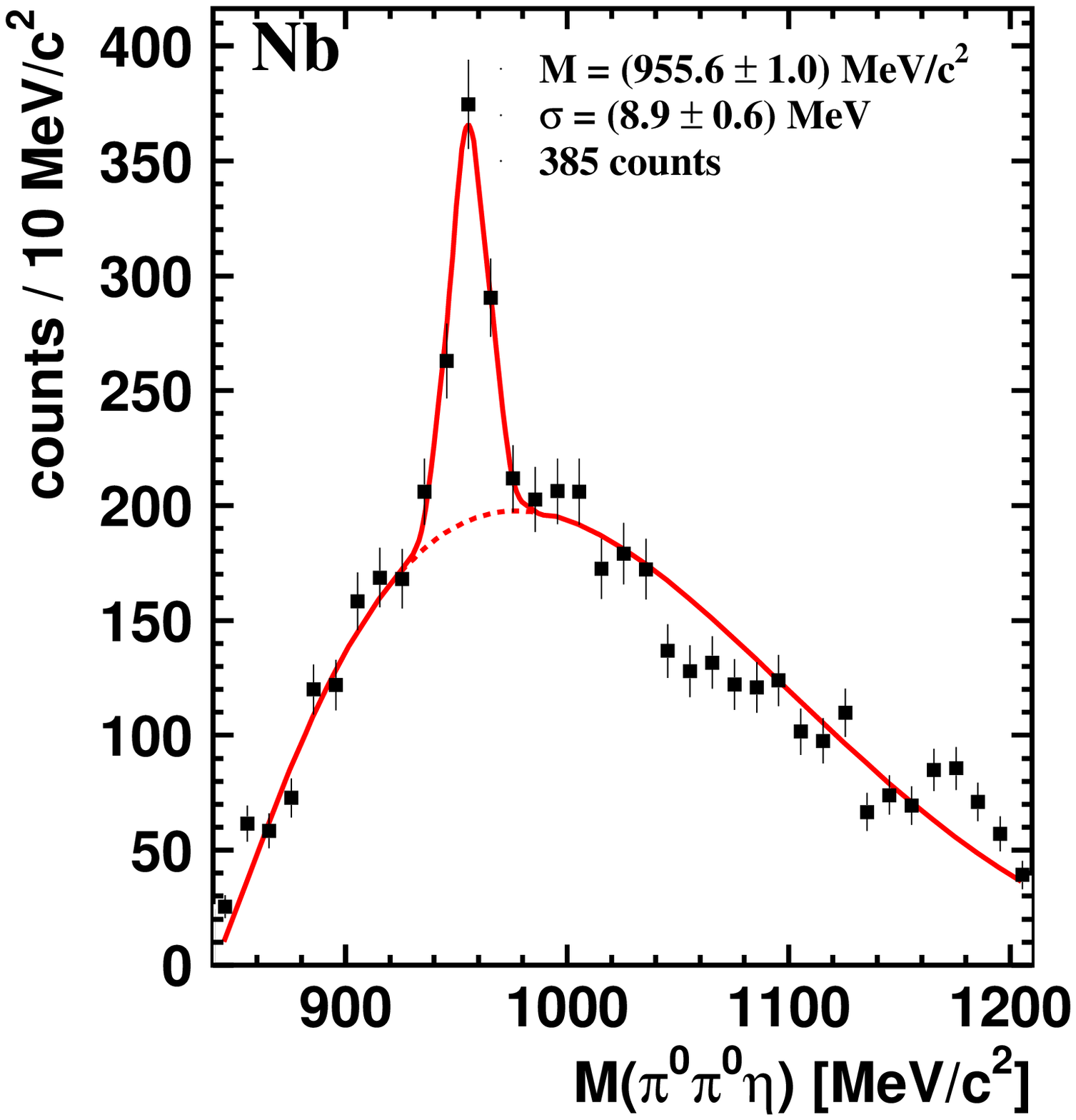} \includegraphics[height=.18\textheight]{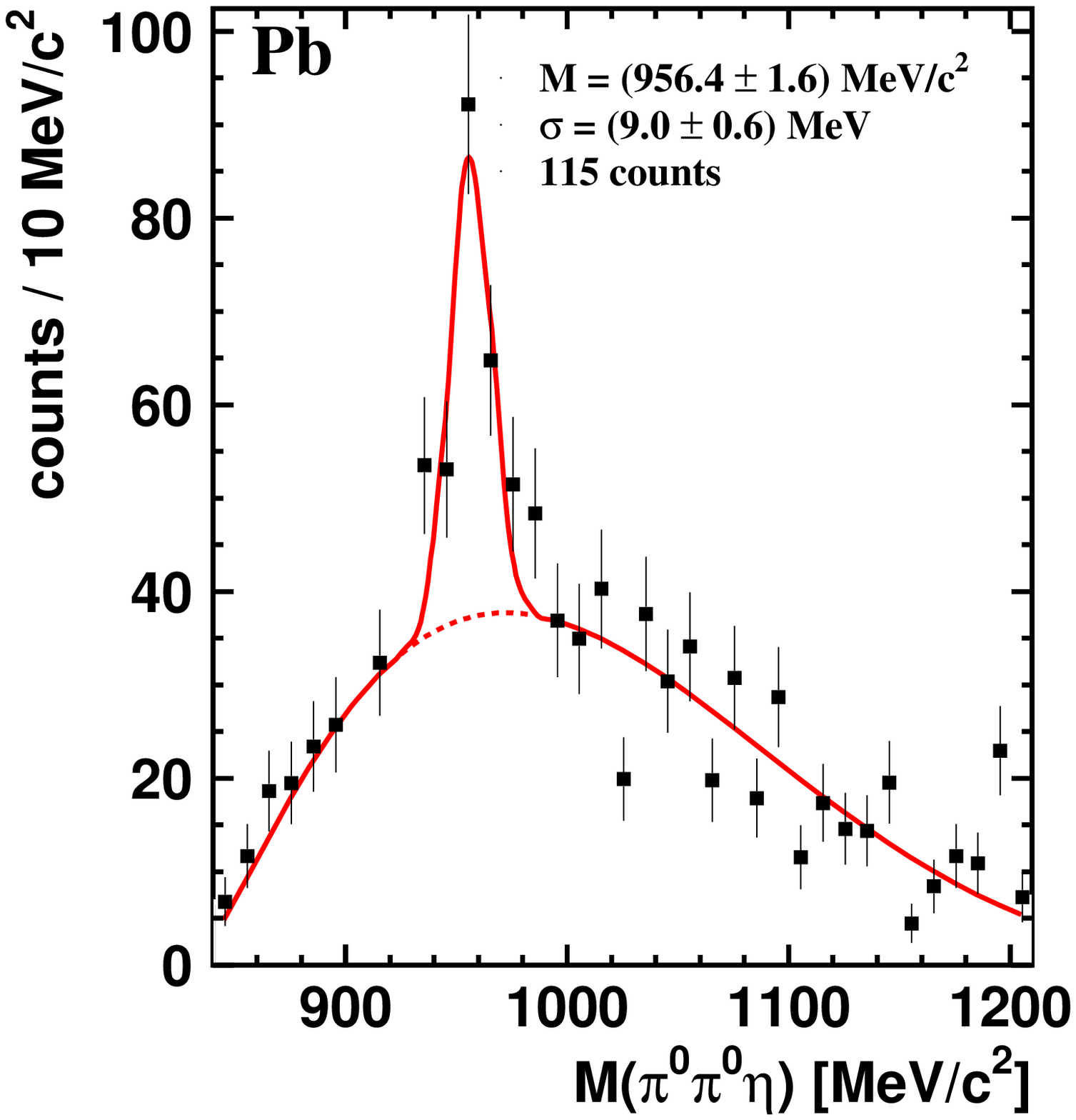}}
 \caption{Invariant mass spectrum of $\pi^{0} \gamma$(upper)~\cite{kotu} and $\pi^{0}\pi^{0}\eta$ (lower)~\cite{nanova} for ${}^{12}\textrm{C}, {}^{40}\textrm{Ca}, {}^{93}\textrm{Nb}$ and ${}^{208}\textrm{Pb}$ targets for the incident photon energy range 900-2200 MeV and 1500 - 2200 MeV respectively. The solid curve is a fit to the spectrum. Only statistical errors are given. See text for more details.} \label{fig:invmass} 
\end{figure*}

The cross section for meson (here $\omega$ or $\eta^\prime$) photoproduction per nucleon within a nucleus is compared with the meson production cross section on a free nucleon $N$. The nucleus serves as a target and also as an absorber. In case of no meson absorption in nuclei this ratio would be one. The transparency ratio is frequently normalized to the transparency ratio measured on a light nucleus like carbon, which helps to suppress the distortion of the transparency ratio by photoabsorption and by two-step processes~\cite{nanova}.

\begin{figure*} 
 \resizebox{0.9\textwidth}{!}{
     \includegraphics[height=.5\textheight]{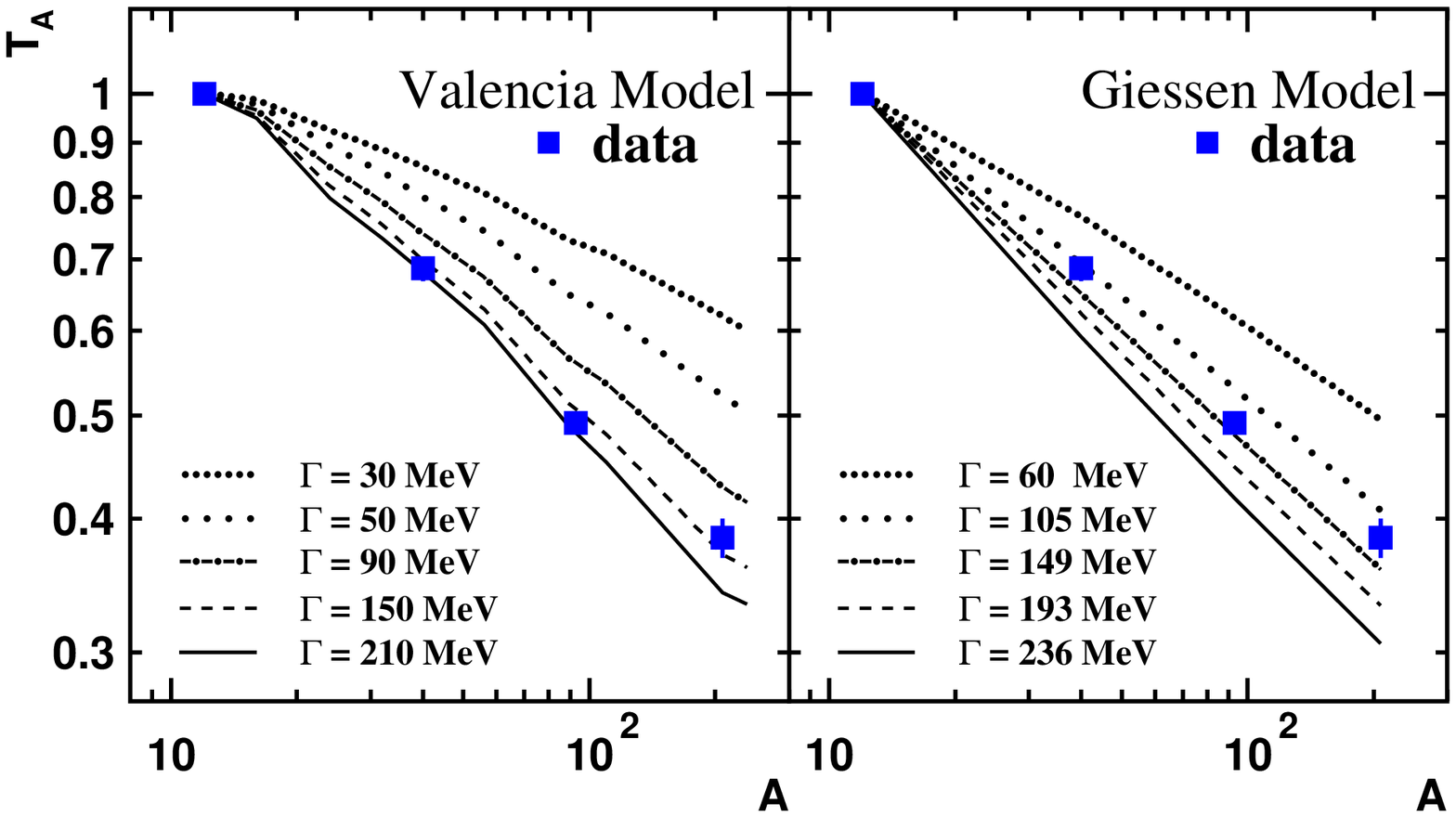} \includegraphics[height=.5\textheight]{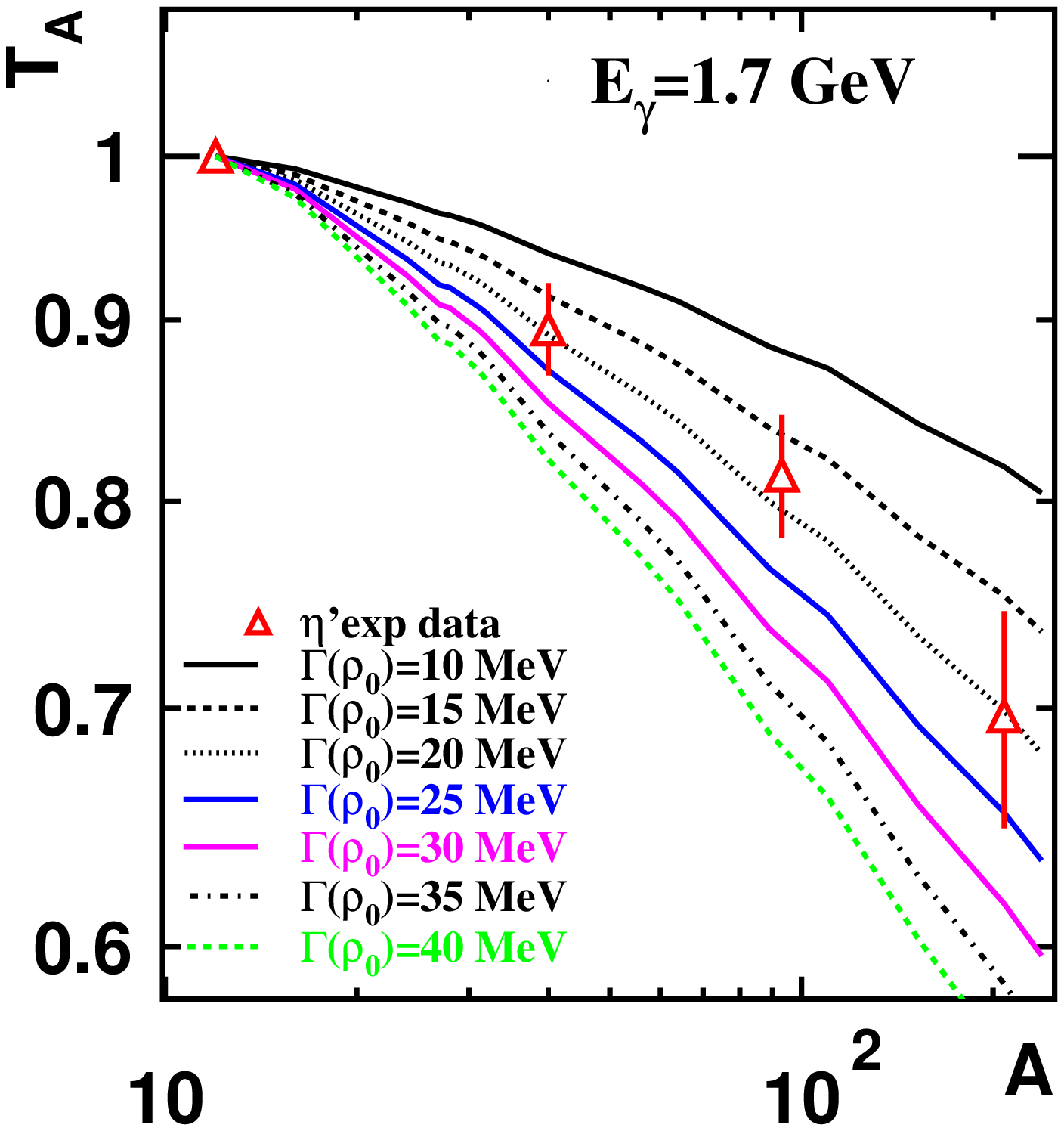}}
 \caption{Transparency ratio of $ \omega$(left and middle)~\cite{kotu} and $\eta^\prime$ (right)~\cite{nanova} relative to that of ${}^{12}\textrm{C}$, as a function of the nuclear mass number $A$. The data are compared with Monte Carlo simulations (left)~\cite{kaskulov}, with BUU transport calculations(middle)~\cite{muhlich1}  and calculations~\cite{nanova} for different values of the in-medium $\eta^\prime$ widths.} \label{fig:ta} 
\end{figure*}

The $\pi^{0} \gamma$ and $\pi^{0} \pi^{0} \eta$ invariant mass distributions corresponding to the $\omega \rightarrow \pi^{0} \gamma$ and $\eta^\prime \rightarrow \pi^{0} \pi^{0} \eta$ decay channels are shown in Fig.~\ref{fig:invmass} for the different solid targets. The spectra were fitted with a Gaussian and a background function $f(m) = a \cdot (m-m_1)^{b} \cdot (m-m_2)^{c} $.  Alternatively, the background shape was also fitted with a polynomial. The measurements have been done in photoproduction experiments on four solid targets (${}^{12}\textrm{C}, {}^{40}\textrm{Ca}, {}^{93}\textrm{Nb}$ and ${}^{208}\textrm{Pb}$) with the Crystal Barrel and TAPS detector system at the ELSA facility in Bonn. The resulting cross sections are used to calculate the transparency ratio of the $\omega$ and $\eta^\prime$-meson for a given nucleus $A$ from the formula~(\ref{eq:trans}), normalized to the carbon data. The results are shown in Fig.~\ref{fig:ta}. The $\omega$ yield per nucleon within a nucleus decreases with increasing nuclear mass number and drops to about 40\% for $Pb$ relative to carbon. A comparison of the data with calculations of the Valencia \cite{kaskulov} and Giessen \cite{muhlich1} theory groups indicates an in-medium width of the $\omega$ meson of about 130-150 $MeV$ at normal nuclear matter density for an average $\omega$ momentum of 1.1 $GeV/c$.  For the $\eta^\prime$ meson an in-medium width of 15-25 $MeV$ is obtained at an average recoil momentum $p_{\eta^\prime} = 1.05\ GeV/c$ ~\cite{nanova}.

\begin{figure}[h!]
 \resizebox{1.\textwidth}{!}{
    \includegraphics[height=0.5\textheight]{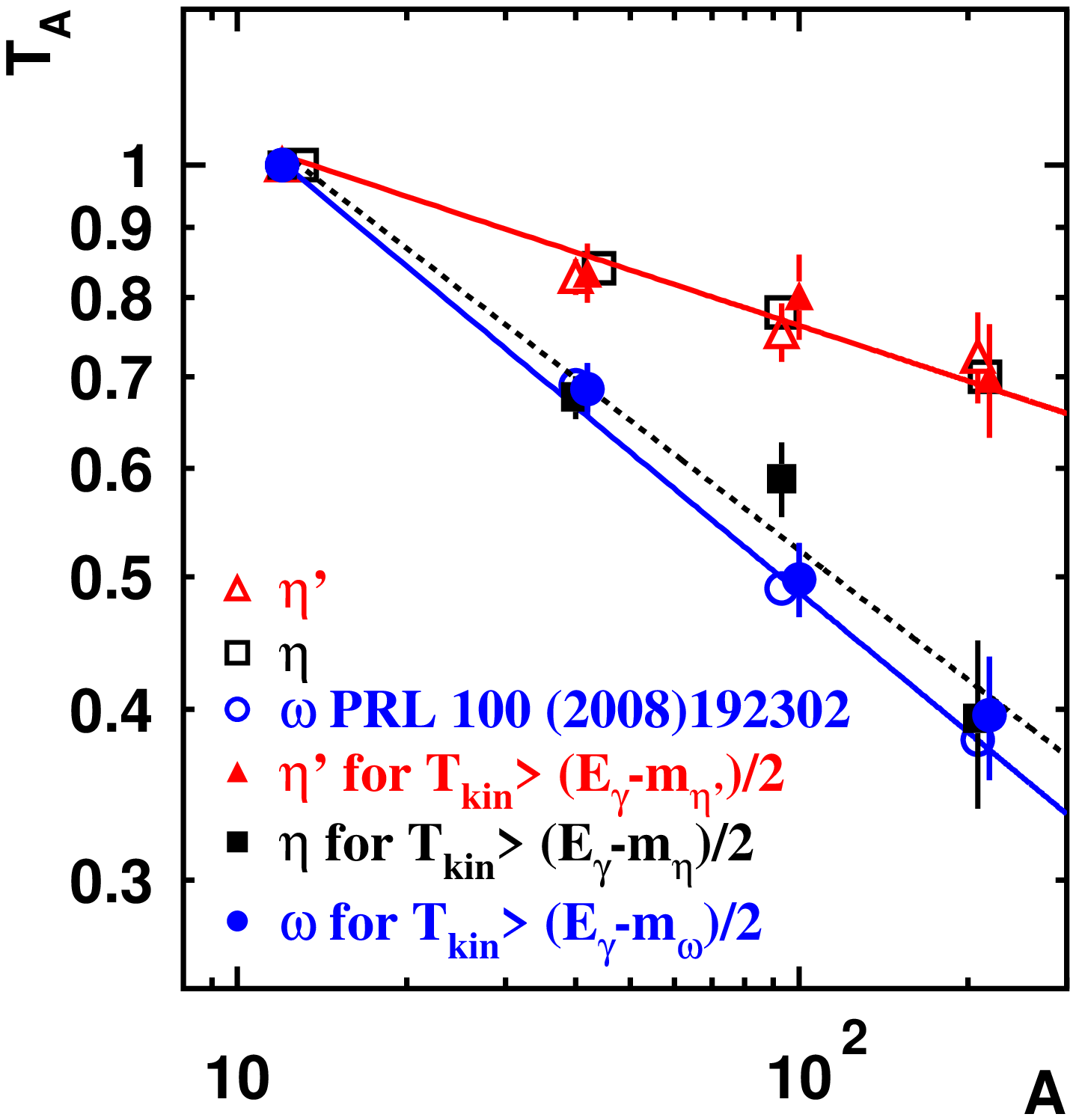}\includegraphics[height=0.5\textheight]{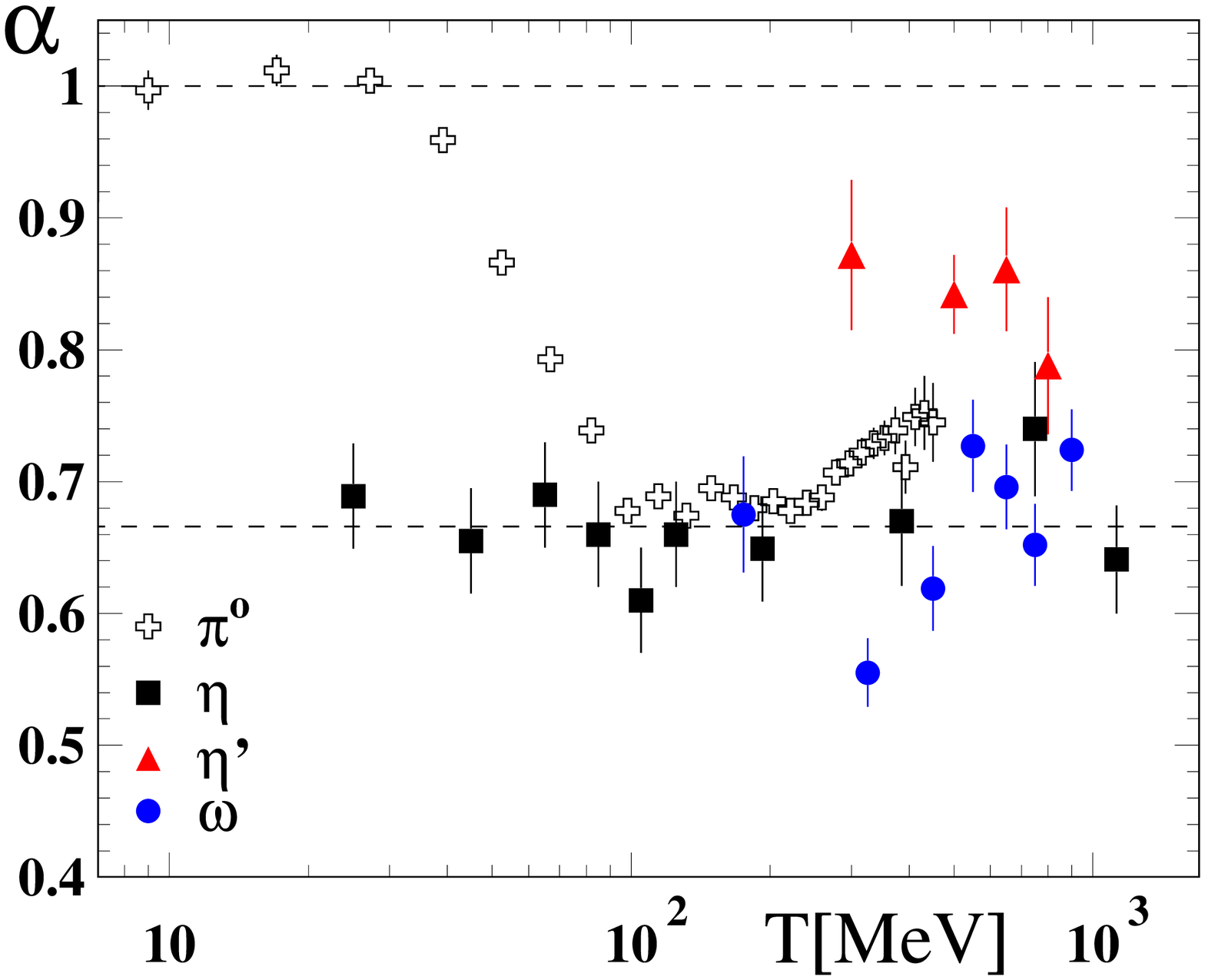}}
\caption{(Left) Transparency ratio for different mesons - $\eta$(squares), $\eta^\prime$(triangels) and $\omega$(circles) as a function of the nuclear mass number $A$.  The transparency ratio with a cut on the kinetic energy for the respective mesons is shown with full symbols. The incident photon energy is in the range 1500 to 2200~MeV. The solid lines are fits to the data. Only statistical errors are shown. The impact of photon shadowing on the determination of the transparency ratio has been taken into account for the $\eta^\prime$ meson, but has not been corrected for in the published data for the other mesons. (Right) $\alpha$ parameter dependence on the kinetic energy T of the meson compared for $\pi^0$ ~\cite{bernd}, $\eta$ ~\cite{thierry, robig}, $\eta^\prime$ and $\omega$ (\cite{kotu}, this work). This figure is an updated version of a figure taken from \cite{thierry}. } 
 \label{fig:taall}
\end{figure}

The transparency ratio extracted as defined in Eq.~(\ref{eq:trans}) is shown in Fig.~\ref{fig:taall}(left) for the $\eta^\prime$-meson (red triangles) compared to the transparency ratio of the $\omega$-meson (blue circles) measured by~\cite{kotu} and the transparency ratio of the $\eta$-meson (black squares) deduced from ~\cite{thierry}. The solid lines are fits to the data points, yielding slope parameters of -0.14 and -0.34 for $\eta^\prime$, 0.34 for $\omega$, respectively.   As it can be seen from the figure, $\eta^\prime$-mesons are only weakly absorbed via inelastic channels as compared to the $\omega$-meson. 

The contribution of secondary production processes like $\pi N \rightarrow  \eta^{'} N$ could increase the number of observed mesons and thus distort the transparency ratio measurement. The mesons reproduced via such processes should have relatively low kinetic energy. Applying a cut on the kinetic energy of the meson it is possible to study the effect of the secondary production processes on the transparency ratio. In Fig.~\ref{fig:taall} the data are shown for the full kinetic energy range  of recoiling mesons (open symbols) as well as for the fraction of high energy mesons (full symbols) selected by the constraint

\begin{equation}
T_{kin} \ge (E_{\gamma} -m)/2. \label{cut}
\end{equation}
 Here, $E_{\gamma} $ is the incoming photon energy and $T_{kin} $ and $m$ are the kinetic energy and the mass of the meson, respectively. As discussed in \cite{thierry}, this cut suppresses meson production in secondary reactions by selecting mesons with higher kinetic energy. Fig.~\ref{fig:taall} shows that within errors this cut does not change the experimentally observed transparency ratios for the $\omega$-meson and $\eta^\prime$-meson while there is a significant difference for the $\eta$-meson. For the latter,  the transparency ratio has changed dramatically and shows a slope of -0.32 (dashed line in Fig.~\ref{fig:taall}) which is quite different from the previous one (slope -0.14 - the empty black squares in Fig.~\ref{fig:taall}).
 For the $\eta$-meson secondary production processes appear to be more likely in the relevant photon energy range because of the larger available phase space due to its lower mass (547~MeV/c$^2$) compared to the $\omega$ (782~MeV/c$^2$) and $\eta^\prime$ (958~MeV/c$^2$) meson. Cross sections for pion-induced reactions favor secondary production processes in case of the $\eta$-meson: 3 mb at $p_{\pi} \approx$ 750 MeV/c in comparison to 2.5 mb at $p_{\pi} \approx $ 1.3 GeV/c for the $\omega$-meson and 0.1 mb at $p_{\pi} \approx$ 1.5 GeV/c for the $\eta^\prime$-meson, respectively \cite{Landolt-Boernstein}. In addition, $\eta$-mesons may be slowed down by absorption of an initially produced $\eta$-meson which is then re-emitted with lower kinetic energy from a S$_{11}$(1535) resonance. According to Fig.~\ref{fig:taall}  the $\eta^\prime$-meson shows a much weaker attenuation in normal nuclear matter than the $\omega$ and $\eta$-meson which exhibit a similarly strong absorption after suppressing secondary production effects in case of the $\eta$-meson.\\
An equivalent representation of the data can be given by parameterizing the observed meson production cross sections by $\sigma(A) = \sigma_0 A^{\alpha(T)}$ where $\sigma_0$ is the photoproduction cross section on the free nucleon and $\alpha$ is a parameter depending on the meson and its kinetic energy. The value of $\alpha \approx$ 1 implies no absorption while $\alpha \approx $ 2/3 indicates meson emission only from the nuclear surface and thus implies strong absorption. All results are summarized in Fig.~\ref{fig:taall} (right) and additionally compared to data for pions \cite{bernd}. For low-energy pions, $\alpha \approx 1.0$ because of a compensation of the repulsive s-wave interaction by the attractive p-wave  $\pi N$ interaction. This value drops to $\approx$ 2/3 for the $\Delta$ excitation range and slightly increases for higher kinetic energies. After suppressing secondary production processes by the cut (Eq.(\ref{cut})) the $\alpha$ parameter for the $\eta$-meson is close to 2/3 for all kinetic energies, indicating strong absorption \cite{thierry}. For the $\omega$-meson the $\alpha$ values are also close to 2/3. The weaker interaction of the $\eta^\prime$-meson with nuclear matter is quantified by $\alpha = 0.84\pm 0.03$ averaged over all kinetic energies.

\subsection{\it Lineshape Analysis}
\begin{figure}[h!]
 \resizebox{1.\textwidth}{!}{
    \includegraphics[height=0.5\textheight]{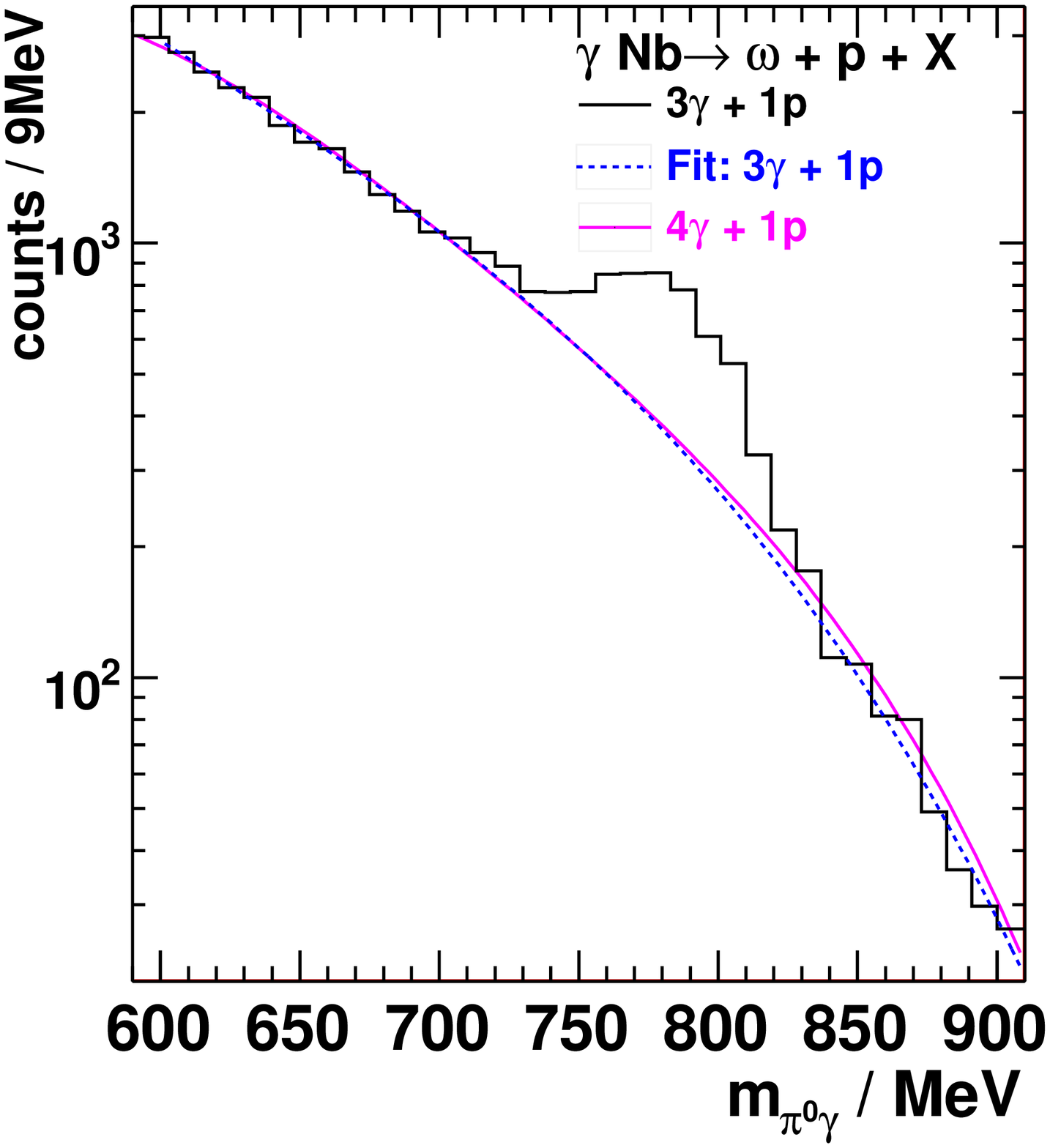}\includegraphics[height=0.5\textheight]{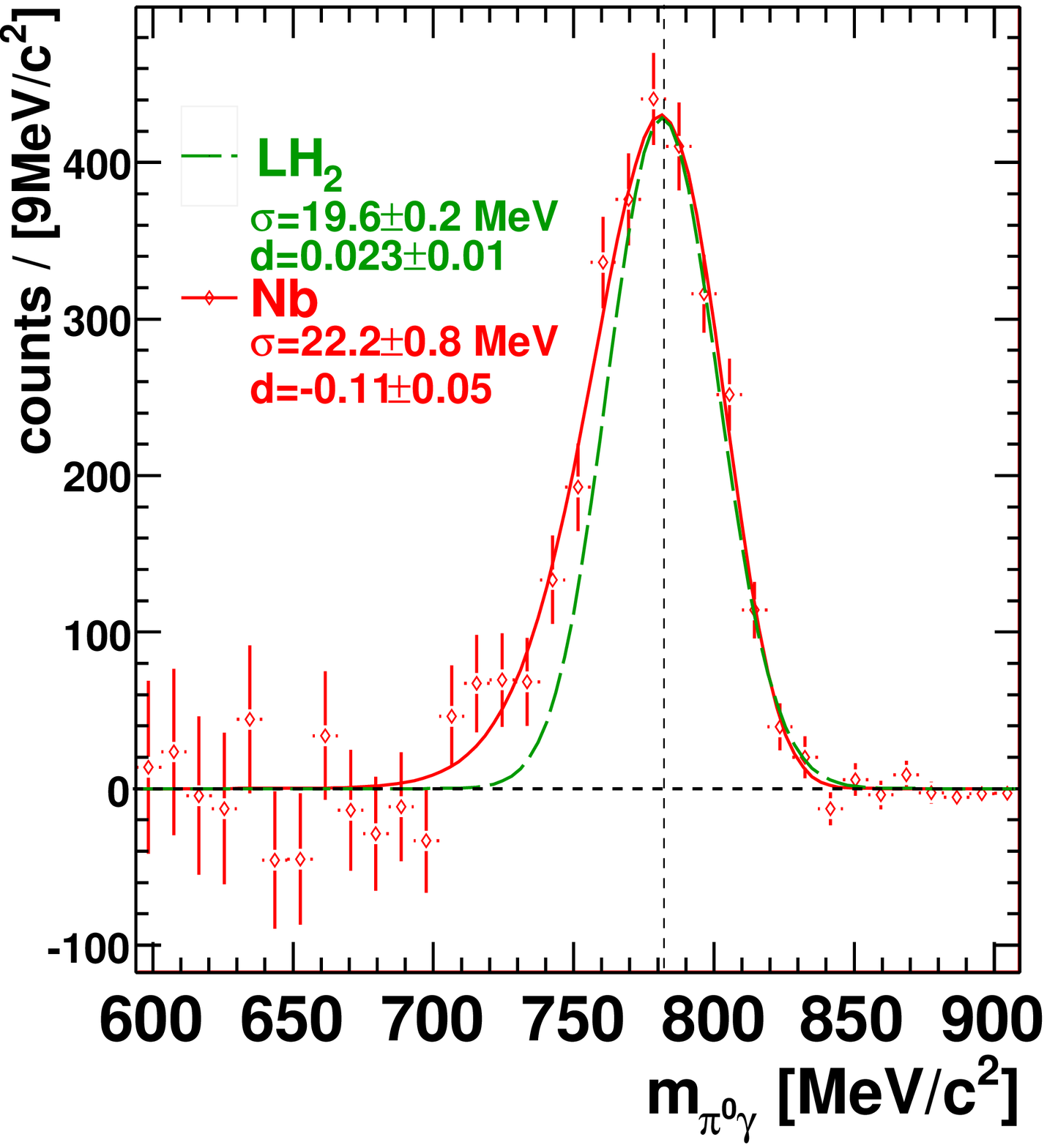} \includegraphics[height=0.5\textheight]{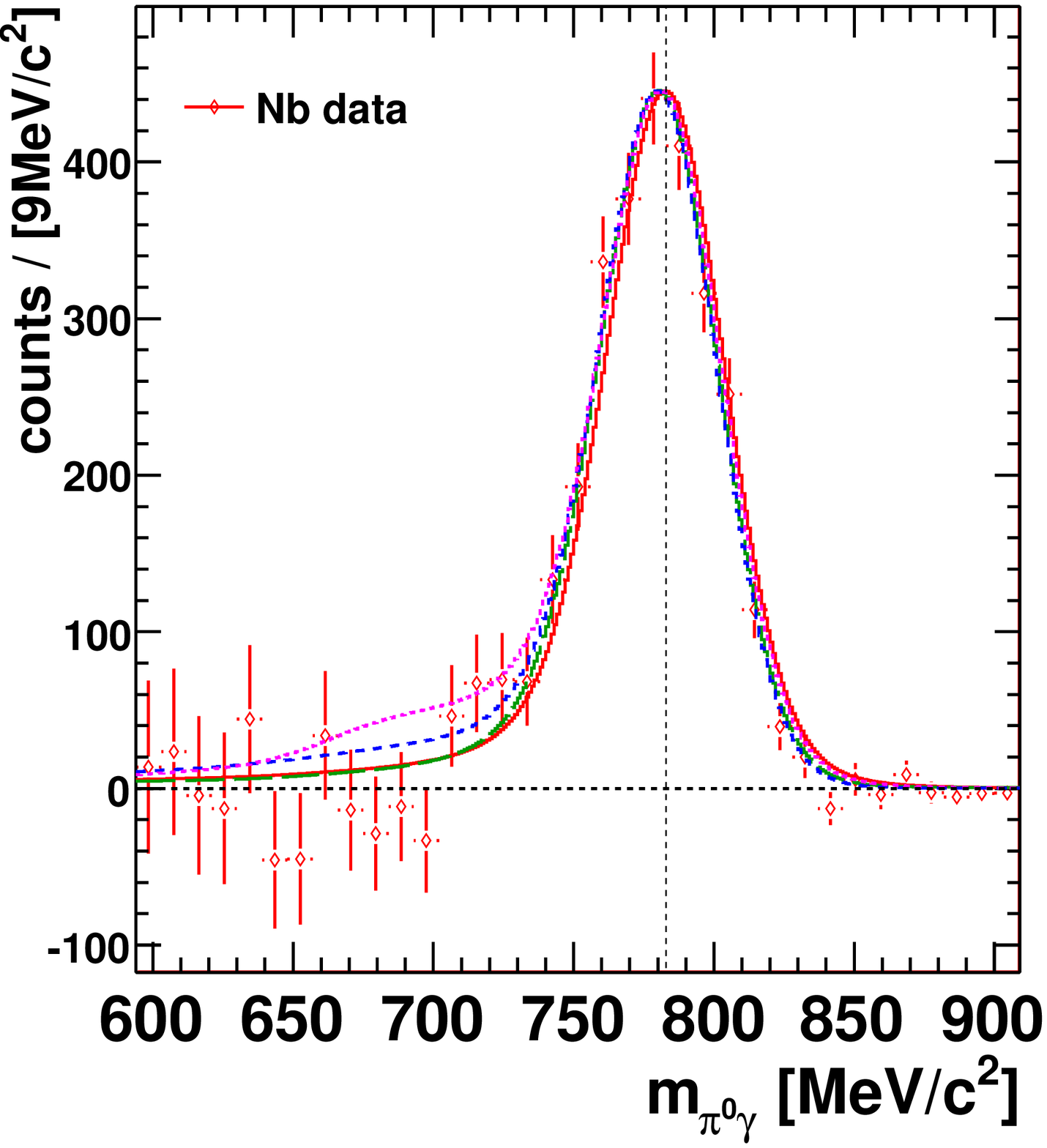}}
\caption{Left: Invariant mass spectrum $\pi^{0} \gamma$ pairs for photons of 900-1300 MeV incident on a $Nb$ target for events with 3 photons and 1 proton registered in the Crystal Ball/TAPS detector system. The dotted (blue) curve is a fit to the background. The solid (red) curve represents the background determined from events with 4 photons and 1 proton, following the procedure described in~\cite{nanova2}. Middle: $\pi^{0} \gamma$ invariant mass spectrum after subtracting the different background shapes. The data points represent the average of the resulting spectra after background subtracting from the two methods (see the text for more details). The dashed (green) curve is a fit to the $\pi^{0} \gamma$ invariant mass spectrum off $LH_{2}$. Right: Comparison of the observed $\omega$ lineshape with GiBUU calculations~\cite{weil, nanova1, metag} for different in-medium modification scenarios, assuming no in-medium modifications (red solid curve), only collisional broadening (green dashed curve), collisional broadening and mass shift by -14\% at normal nuclear matter density (short dashed, blue curve) and mass shift without broadening (dotted, magenta curve).}
 \label{fig:micha}
\end{figure}

The mass $\mu$ of the meson can be deduced from the 4-momentum vectors $p_{1}, p_{2}$ of the decay products according to
\begin{equation}
\mu(\vec p, \rho) = \sqrt{(p_{1}+p_{2})^{2}}. \label{mass}
\end{equation}
The mass distribution $\mu(\vec p,\rho)$ depends on the 3-momentum $\vec p$ of the vector meson and on the density $\rho$ of the nuclear medium at the decay point. Only mesons decaying inside the nucleus carry information on in-medium properties which are to be studied. An in-medium mass shift of the meson could be observed by comparing the mass calculated from Eq.~(\ref{mass}) in the limit of low meson momenta with the vacuum mass of this meson listed in~\cite{nakamura}.
The light vector mesons $\rho, \omega$ and $\Phi$ are particulary suited for the mass distribution measurements since their lifetimes of 1.3 $fm/c$, 23 $fm/c$ and 46 $fm/c$, respectively, are so short that they decay within the nuclear medium with some probability after production in a nuclear reaction. Nevertheless, momentum cuts have to be applied for the longer lived $\omega$ and $\Phi$ mesons to achieve decay lengths comparable to nuclear dimensions. Obviously the line shape analysis is not applicable for $\eta^\prime$ meson in the medium since its decay length is much larger than the nuclear dimension.
Calculating the in-medium mass of the meson from Eq.~(\ref{mass}) requires that the 4-momentum vectors of the decay products are not distored by final state interactions. For this reason dileptons are prefered decay channel because any final state distortion of the 4-momenta of the decay products is avoided. Unfortunately, the branching ratios into the dilepton channels are only of the order of 10$^{-5}$ - 10$^{-4}$, making these measurements very difficult. Here, we present our investigations of in-medium modifications of the $\omega$-meson in $\omega \rightarrow \pi^{0} \gamma$ decay channel with a branching ratio of 8.9\%.  As shown in \cite{johan}, events with a final state interaction of the $\pi^{0}$ meson can be sufficiently suppressed by applying a cut on the kinetic energy of the pions $T_{\pi} \ge 150 \ MeV$.
As pointed out in~\cite{leopold}, the measured mass distribution in experiments with elementary probes represents a
convolution of the hadron spectral function with the branching ratio
$\Gamma_{H \rightarrow X_{1}X_{2}}/\Gamma_{tot}(m)$ into the channel being studied.  Since
this branching ratio depends on the invariant mass $m$ this
may lead to deviations of the experimentally determined mass distribution from
the true spectral function. In addition it is demonstrated that for a strong
in-medium broadening of the hadron - as observed for the $\omega$-meson
~\cite{kotu} -  contributions from higher densities are suppressed, thereby reducing the sensitivity of the $\omega$ signal shape to in-medium modifications. Furthermore, ambiguities in the subtraction of the background in $\pi^{0} \gamma$ invariant mass spectra leads to additional uncertainties in the determination of the $\omega$ lineshape, as discussed in \cite{kaskulov1, david, nanova2}. In a series of experiments with tagged photon beams at MAMI-C, using the Crystal Ball and TAPS detector system, $\omega$ photoproduction on ${}^{12}\textrm{C}$ and ${}^{93}\textrm{Nb}$ targets has been measured in the energy range close to the production threshold 900-1300 MeV~\cite{thiel}.
The $\pi^{0} \gamma$ invariant mass spectrum obtained on $Nb$ is shown in Fig.~\ref{fig:micha}(left).
The background under the $\omega$ peak is mostly comming from 2$\pi^{0}$ and $\pi^{0} \eta$ events where one of the 4 decay photons was not registered in the Crystal Ball/TAPS detector system. The background has been determined by two methods: 1) a fit with a 5th order polynomial and 2) a method developed and described in~\cite{nanova2} where the background is derived from the measured  2$\pi^{0}$ and $\pi^{0} \eta$ events. Fig.~\ref{fig:micha} (middle) shows the average of $\omega$ signal after subtraction of the different background shapes. The $\omega$ signal is compared to the $\omega$ signal for a $LH_{2}$ target, which is used as a reference measurement. The signal on $Nb$ is slightly broader than the reference signal obtained with the $LH_{2}$ target, consistent with the in-medium broadening of the $\omega$ meson found in the transparency ratio measurement. In Fig.~\ref{fig:micha}(right) the $\omega$ signal on Nb is compared to GiBUU transport calculations~\cite{nanova1,weil} for different in-medium scenarios. Although the statistics of the experiment has been considerably improved compared to previous studies~\cite{david, nanova2, nanova1} it is still not sufficient to distinguish different scenarios which differ only very little in the predicted lineshape. The data disfavore the scenario with mass shift.

\subsection{\it Momentum Distribution Analysis}
\begin{figure*}
  \begin{center}
 \resizebox{0.7\textwidth}{!}{
    \includegraphics[height=0.5\textheight]{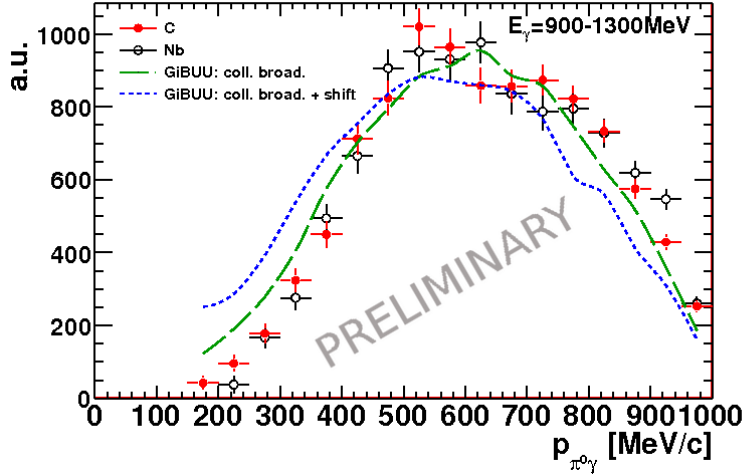}}
\caption{Momentum distribution of $\omega$ mesons for photons of 900 - 1300 MeV incident on $C$ (open black circles) and $Nb$ (closed red circles) targets. The data are compared to GiBUU transport model calculations~\cite{weil}, assuming collisional broadening of the $\omega$ meson with and without an in-medium mass shift of -14\% at normal nuclear matter density, respectively.}
 \label{fig:mom}
 \end{center}
\end{figure*}

If the mass of a meson produced in a nuclear environment were lowered, the difference to the nominal meson mass would have to be compensated at the expense of its kinetic energy upon leaving the nucleus. As demonstarted in GiBUU transport model calculations~\cite{weil}, this leads to a downward shift in the momentum distribution as compared to a scenario without mass shift. A mass shift can thus be indirectly inferred from a measurement of the momentum distribution of the meson.
The momentum distribution for $\omega$ mesons for incident photon energies between 900 and 1300 MeV is shown in Fig.~\ref{fig:mom}~\cite{metag}. The preliminary data points for $C$ and $Nb$ do not show any significant difference. The comparison with GiBUU calculations~\cite{weil, metag} clearly favors the collisional broadening scenario without mass shift.

% \begin{figure}[h!]

\subsection{\it Excitation Function Measurement} 
\begin{figure*}
  \begin{center}
 \resizebox{0.7\textwidth}{!}{
    \includegraphics[height=0.5\textheight]{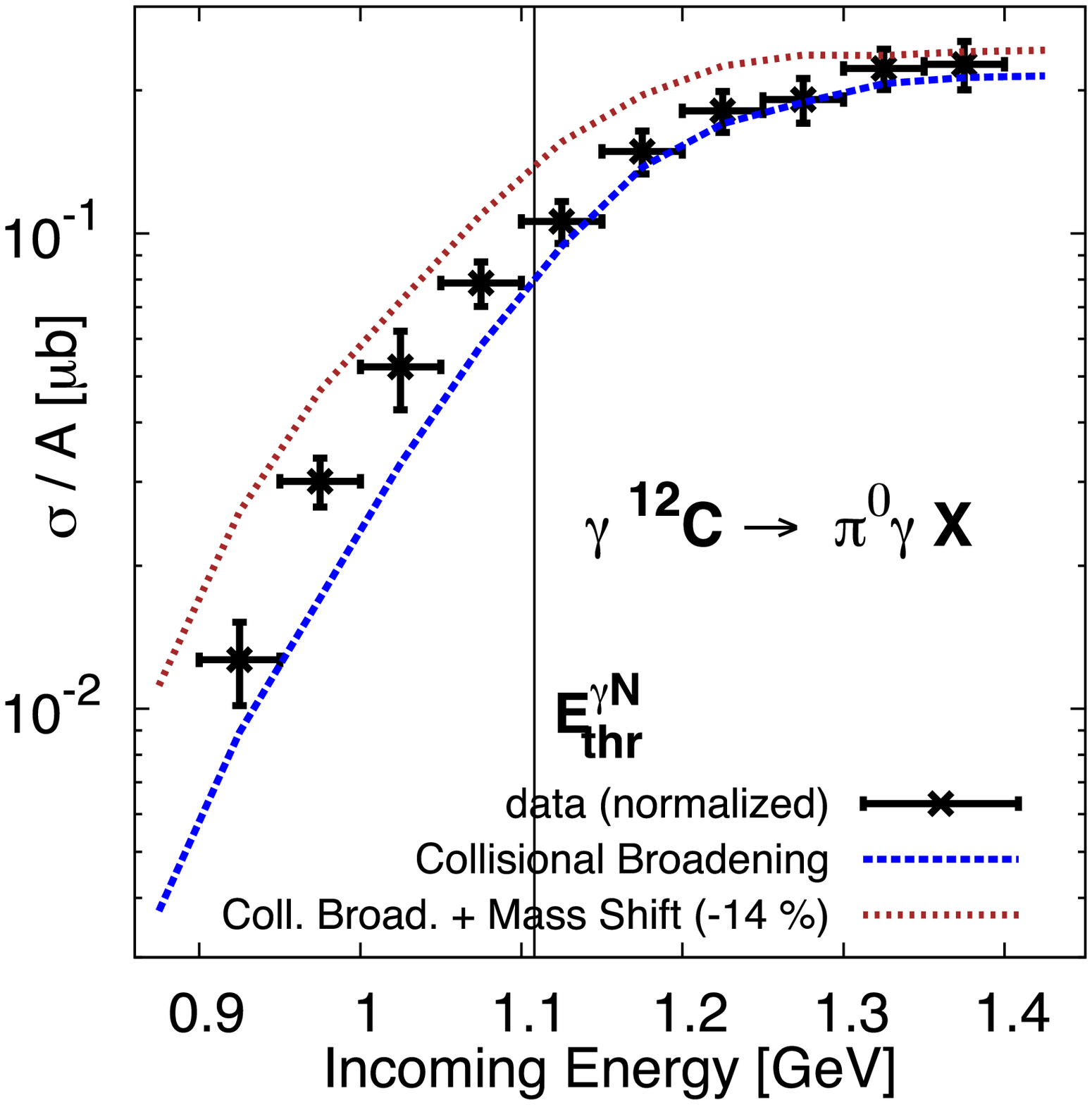} \  \includegraphics[height=0.5\textheight]{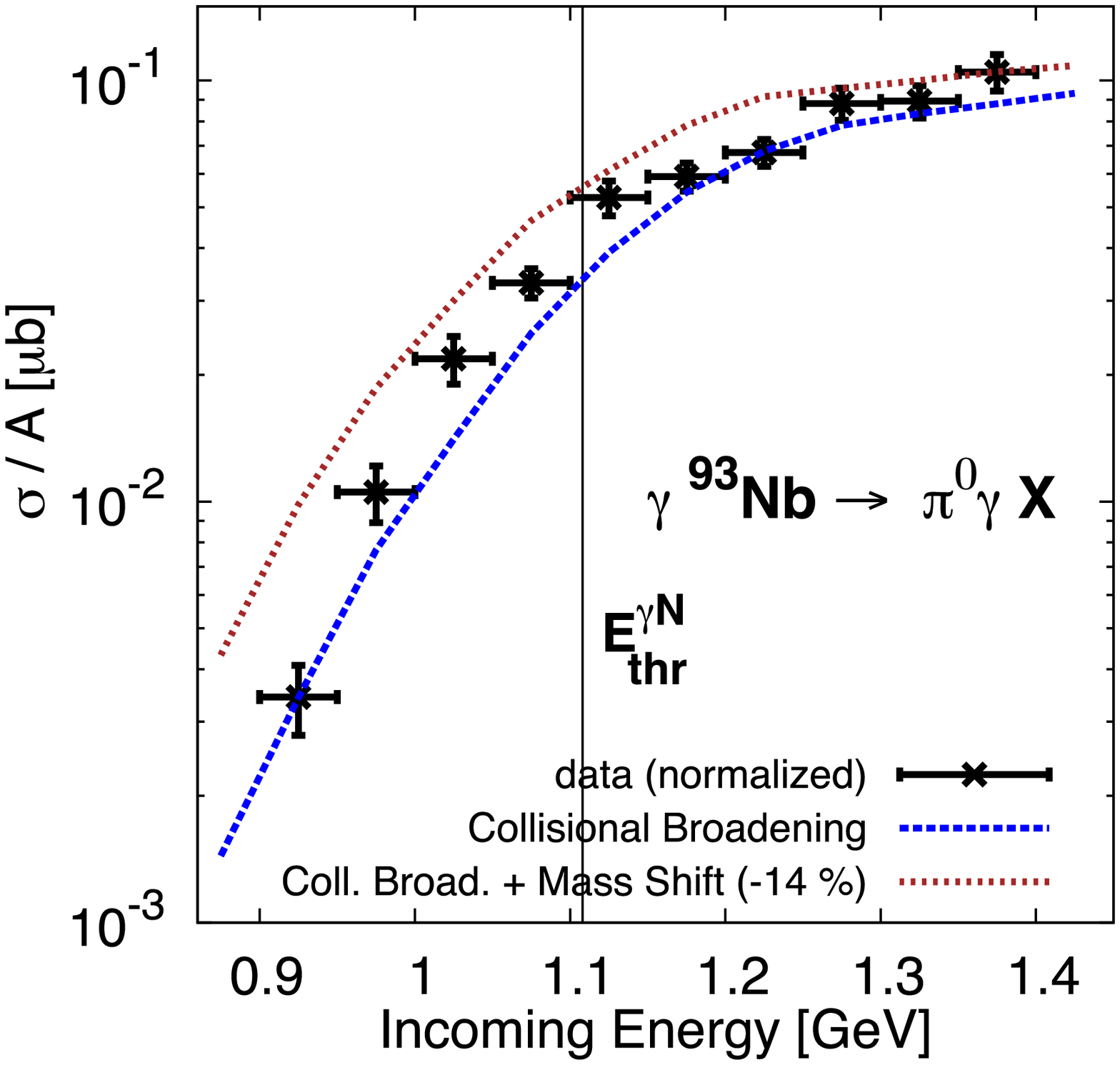}}
\caption{Excitation functions for photoproduction of $\omega$ mesons off $C$ (left) and $Nb$ (right), respectively. The vertical lines indicate the threshold energy for $\omega$ photoproduction on a free nucleon. The experimental data are compared to GiBUU transport model calculations~\cite{metag, weil}, assuming collisional broadening of the $\omega$ meson with and without an in-medium mass shift of -14\% at normal nuclear matter density, respectively. The data are normalized to the calculations at an incident photon  energy of 1375 MeV.}
 \label{fig:excit}
 \end{center}
\end{figure*}

Due to a possible mass shift  the energy threshold for producing the meson is lowered leading to an increase in phase space. As a consequence the production cross section of the meson near the threshold will increase in case of an in-medium mass shift as compared to a scenario without mass shift~\cite{muhlich1}. The experimental approach to study the in-medium properties of the meson and possible mass shift is to measure the excitation function near the production threshold of the meson on different nuclear targets. The experimental data on the cross section for $\omega$ photoproduction on $C$ and $Nb$ are shown in Fig.~\ref{fig:excit}. The data are taken with the Crystal Ball/TAPS detector system at the MAMI-C accelerator in Mainz in the photon energy range 900-1400 MeV. The partial cross section for the $\omega \rightarrow \pi^{0} \gamma$ decay is plotted as a function of the photon incident energy in comparison with GiBUU simulations for the collisional broadening scenario with and without mass shift~\cite{metag}. Since the absolute normalisation of the experimental cross sections has not yet been determined both data sets have been normalized to the respective calculated cross sections at an incident photon energy of 1375 MeV. Again the scenario without mass shift is favored. For a given incident photon energy larger cross sections would have been expected in case of a dropping in-medium mass of the $\omega$ meson. A smaller mass shift, however, can not be excluded from this comparison.
 
% \begin{figure}[h!]

 \section{Conclusion}
 The experimental approaches to study the in-medium properties of mesons are presented and discussed. The transparency ratios for $\eta^\prime$-mesons have been measured for several nuclei. A comparison of these results with corresponding measurements for $\omega$- and $\eta$-mesons demonstrates the relatively weak interaction of the $\eta^\prime$-meson with nuclear matter. The in-medium width of the $\eta^\prime$-meson is found to be 15 - 25 MeV. This relatively small width encourages the search for $\eta^\prime$-nuclear bound states~\cite{Nagahiro}.
 Even with improved statistics an analysis of the $\omega$ meson lineshape does not allow a discrimination between different in-medium modification scenarios. Indirect approaches like analysis of the momentum distribution and the excitation function of $\omega$ mesons seem to favor a scenario of collisional broadening without mass shift. \\

\section{Acknowledgment}
I would like to thank V. Metag and the PhD students and postdocs of our group, in particular H. Bergh\"auser, S. Friedrich, B. Lemmer, K. Makonyi and M. Thiel for their contribution to this talk.\\
This work was supported by  DFG through SFB/TR16.

\end{document}